\documentclass[aps,
preprint,
amsmath,amssymb,
aps,
prfluids]{revtex4-2}

\usepackage[svgnames]{xcolor}
\usepackage{float}
\usepackage{placeins}
\newcommand{\red}[1]{\textcolor{black}{#1}}
\usepackage{graphicx}
\usepackage{dcolumn}
\usepackage{bm}
\usepackage{hyperref}
\usepackage[mathlines]{lineno}

\begin{document}
\preprint{Submitted to Physical Review Fluids}



\title{How thermally-induced secondary motions in offshore hybrid wind--solar farms improve wind-farm efficiency}%

\author{Thijs Bon}%
\email[]{thijs.bon@kuleuven.be}
\affiliation{%
	KU Leuven, Mechanical Engineering, Celestijnenlaan 300, B3001 Leuven, Belgium
}%

\author{Vincent Van Craenenbrock}
\affiliation{%
	KU Leuven, Mechanical Engineering, Celestijnenlaan 300, B3001 Leuven, Belgium
}%

\author{Johan Meyers}%
\affiliation{%
	KU Leuven, Mechanical Engineering, Celestijnenlaan 300, B3001 Leuven, Belgium
}%


\date{\today}

\begin{abstract}
Integrating floating photovoltaic (FPV) installations into offshore wind farms has been proposed as a major opportunity to scale up offshore renewable energy generation. The interaction between these hybrid wind--solar farms and the atmospheric boundary layer (ABL) is for the first time addressed in the present study. Idealized large-eddy simulations (LES) are used to investigate the flow through both isolated wind and solar farms, as well as combined wind--solar farms, in varying configurations and under different atmospheric conditions. When the FPV modules are arranged in long strips parallel to the flow direction, secondary motions arise due to the temperature difference between the warm FPV modules and colder sea surface, significantly affecting the horizontal distribution of mean wind speed in the ABL. Associated downdrafts between the strips increase entrainment of high-speed momentum from above, thereby increasing the wind speed at these locations. LES of hybrid wind--solar farms reveals that this is beneficial for wind turbines when they are located between the FPV arrays, leading, for the cases that we considered, to a farm-averaged power increase of up to 30\% compared to an isolated wind farm. It is shown that the ratio between inertial and buoyancy forces, quantified by a heterogeneity Richardson number $Ri_h$, plays a crucial role in the formation of secondary flows and the resulting wind-farm power enhancement. We further show that no significant power gains, but also no losses emerge in situations when the solar panels are aligned perpendicular to the flow. Although the present results suggest a large potential for wind--solar hybridization, future research should focus on a wider range of scenarios, obtained from a realistic distribution of wind directions and speeds to quantify the potential beneficial impact on real annual power production.
\end{abstract}


\maketitle


\section{Introduction}
As global demand for sustainable energy surges and available land for new infrastructure becomes scarce, technologies for offshore power generation are emerging. While the first offshore wind farm experiments were only started back in the 1990's \citep{Esteban2011}, the present global offshore wind capacity is 64.3 GW (in 2022) and is expected to grow with another 380 GW in the next decade \citep{GlobalWindEnergyCauncil2023}. By contrast, the first solar energy project on open sea was developed not earlier than 2017 \citep{OceansOfEnergy2024}, hence the technology for floating photovoltaic (FPV) systems in marine areas is still in its infancy. 
An alleged advantage of FPV over land-based systems is higher photovoltaic efficiency and durability due to lower operating temperatures, typically attributed to the cooling effect of the underlying water or convective heat transfer to the air, depending on the system design and wind speed (\citet{Micheli2022}, and references therein). The main challenges of offshore FPV installations are related to the harsh conditions, as the structures should withstand high wind speed and waves as well as corrosion from salty water \citep{Ghosh2023}. Nevertheless, energy production at sea is expected to become a major source of growth in the photovoltaic and wind industry, as it circumvents the problem of intensive land use, which is particularly important for densely populated areas such as Western Europe.

With the availability of technologies to generate both solar and wind energy in marine areas, the concept of combined offshore wind--solar farms has been recently explored \citep[e.g.][]{Lopez2020, Golroodbari2021, Solomin2021, Shi2023, Delbeke2023}. In fact, the first marine hybrid FPV--Wind projects have already been initiated in Shandong, China \citep{Shi2023} and on the Dutch North Sea \citep{OceansOfEnergy2024}. Adding FPV systems to large offshore wind farms is an attractive option given the large space available in between wind turbines, as well as the already available or planned cable capacity that connects the wind farm to the grid on land \citep{Golroodbari2021}. Another major advantage is the spatiotemporal complementarity of wind and solar resources, which reduces variability in power output of a hybrid wind--solar farm and significantly increases the energy production per unit of marine area \cite{Lopez2020, DeSouzaNascimento2022, Delbeke2023}.

An effect that remains largely unexplored to the authors' best knowledge, is the influence that large offshore FPV arrays may have on the atmospheric boundary layer (ABL) flow above it. Recent studies have demonstrated that, when the mean flow direction is parallel to long strips of varying aerodynamic or thermal properties, large-scale vortices can be generated \citep[e.g.][and references therein]{Barros2014, Medjnoun2020, Salesky2022, Bon2022}. These surface-induced secondary motions affect the entire ABL and produce `high- and low momentum pathways' (HMPs and LMPs), i.e. time-averaged spatial variations in wind speed. The secondary flows and their effects are most pronounced if the spanwise length scale of the surface heterogeneity is of the same order as the turbulent boundary-layer height \citep[e.g.][]{Medjnoun2018, Bon2022}, i.e. about 100~m to 3~km \citep{Stull1988}. Given that the typical spacing between wind turbines in large offshore wind farms is about 1 km  \cite[e.g.]{Barthelmie2010}, and the size of current FPV solar farms is also reaching the km range \citep{Benjamins2024}, heterogeneity-induced secondary flows can be expected to be relevant for offshore wind--solar farms. Moreover, realizing that the power that a wind turbine can extract from the flow is proportional to the incoming wind speed cubed, relatively small horizontal wind speed variations related to secondary motions ($\sim1$ m s$^{-1}$ in \citet{Salesky2022}) may significantly affect wind farm power output. Therefore, the formation of secondary flows may be an important consideration in determining the location of wind turbines and FPV arrays relative to each other. 

\begin{figure}[h]
	\centering
	\includegraphics[width=\textwidth,trim={0cm 2cm 0cm 1cm},clip]{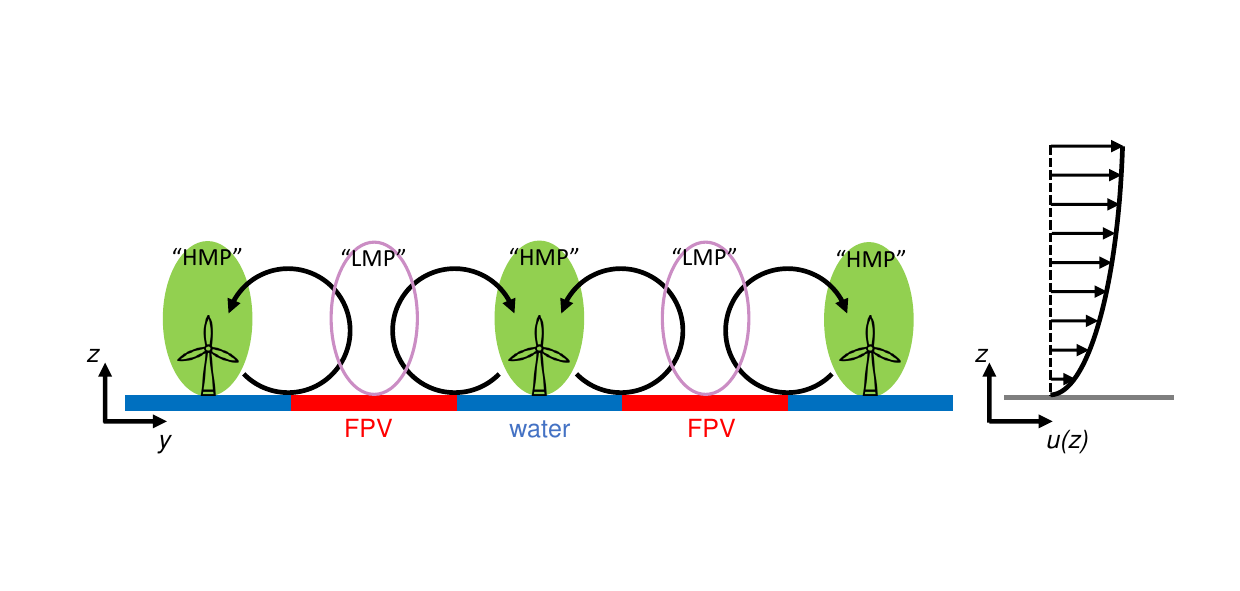}
	\caption{\red{Conceptual picture of how thermally-induced secondary flows over streamwise-aligned FPV arrays would enhance wind speed at turbine locations. Red resembles FPV arrays with high surface temperature and blue sea surface with lower surface temperature. Secondary vortices (black arrows) mix high-speed air from higher up in the boundary layer downwards, resulting in high-momentum pathways (HMPs, green areas). A typical sheared boundary-layer wind speed profile is depicted at the right-hand side.}   }
	\label{fig:schematic}
\end{figure}

This work focuses on thermally induced secondary flows, arising from the expected temperature difference between the sea surface and FPV installations. The formation of secondary motions due to variation in surface temperature, and associated HMPs and LMPs, is illustrated in Fig. \ref{fig:schematic}. Above areas that have higher temperature than their environment, such as FPV arrays surrounded by sea, air rises due to buoyancy forces, mixing slow air from near the surface upwards and thus reducing wind speed. The opposite happens in the surrounding colder areas, where air from greater heights is mixed downwards, thereby increasing mean wind speed in these locations.

The present study aims to (i) examine the significance of secondary flows generated by surface temperature differences associated with large FPV arrays, (ii) identify key parameters affecting this phenomenon, and (iii) explore to what extent the related horizontal velocity variations could be exploited to improve wind farm performance. Concerning the last point, the working hypothesis is that the power output of a wind farm would increase when the turbines are placed within a high-momentum pathway, in between the FPV arrays, as depicted in Fig. \ref{fig:schematic}.

In order to model the atmospheric flow through a hybrid wind--solar farm, we employ idealized large-eddy simulations (LES). This technique has been extensively used to study the interaction between wind farms and the ABL under neutral, stably and unstably stratified flow conditions \citep{Calaf2010, Allaerts2018, Porte-Agel2020, Gadde2021, Maas2022, Lanzilao2024}, as well as the effects of thermal surface heterogeneity \citep{Margairaz2020, Salesky2022, Fogarty2023}. The novelty of the present research lies in the combination of wind farms and non-uniform surface temperature. 
We focus on the neutrally stratified boundary layer, which is a reasonable assumption since the diurnal cycle is weak in offshore conditions. Following e.g. \citet{Calaf2010} and \citet{Salesky2022}, the pressure-driven boundary layer (PDBL) approach is adopted, where Coriolis forces are neglected. 

This manuscript is built up as follows: first, the simulation framework and considered cases are outlined in section \ref{sec:methodology}. In section \ref{subsec:secondaryflows}, we briefly examine secondary motions induced by large FPV arrays for varying FPV temperature and wind speed, without the presence of a wind farm. In section \ref{subsec:sm_wsf} simulations of hybrid wind--solar farms are analysed, including an investigation of the effects of driving pressure and temperature forcing (\ref{subsubsec:WF_Ri}), solar farm layout (\ref{subsubsec:SF_Layout}) and wind direction (\ref{subsubsec:wind_direction}). The impact of idealizing assumptions in the present study are discussed in section \ref{sec:discussion}. Finally,  our conclusions are summarized in section \ref{sec:conclusion}. 

\FloatBarrier





\section{Numerical Methodology}
\label{sec:methodology}
\subsection{Large-eddy simulation method}
The governing equations for the filtered variables in the LES are the continuity equation and Navier--Stokes equations, coupled with a transport equation for potential temperature:

\begin{equation}
	\frac{\partial u_i}{\partial x_i} = 0
	\label{eq:continuity}
\end{equation}
\begin{equation}
	\frac{\partial u_i}{\partial t} + \frac{\partial \left(u_i u_j\right)}{\partial x_j} = -\frac{\partial p^*}{\partial x_i}- \frac{\partial p_{\infty}}{\partial x_i}\delta_{i1}-\frac{\partial \tau^{sgs}_{ij}}{\partial x_j} +  \frac{g}{\theta_0} \left(\theta - \langle \theta \rangle \right)\delta_{i3} + f_i
	\label{eq:NS}
\end{equation}
\begin{equation}
	\frac{\partial \theta}{\partial t} + \frac{\partial \left(u_j\theta\right)}{\partial x_j} = - \frac{\partial q_j^{sgs}}{\partial x_j}
	\label{eq:temperature}
\end{equation}
where $i=1,2$ denote the horizontal streamwise ($x$) and spanwise ($y$) directions, $i=3$ indicates the vertical coordinate and $\delta_{ij}$ is the Kronecker delta function. The velocity components $(u_1,u_2,u_3)=(u,v,w)$ and potential temperature $\theta$ represent the resolved variables, filtered on the LES grid. Buoyancy forces are modelled through the Boussinesq approximation, where $g=9.81$ m s$^{-2}$ is the gravitational acceleration and $\theta_0$ is a reference potential temperature, while $\langle \cdot \rangle$ indicates a horizontally-averaged quantity. The PDBL is forced in the streamwise direction by a constant background pressure gradient $\mathrm{d} p_\infty/\mathrm{d} x$. Furthermore, the modified pressure perturbation $p^* = (p-p_\infty)/\rho_0 + \tau^{sgs}_{kk}/3$ includes the trace of the subgrid-scale stress tensor, where $\rho_0$ is a reference density for dry air. The subgrid-scale stress and heat flux tensors $\tau_{ij}^{sgs}$ and $q_j^{sgs}$, account for the effect of unresolved motions. The last term, $f_i$, accounts for additional external forces exerted on the flow, such as the fringe region forcing (see below) and the turbine thrust force $f_i^{turb}$. 

The governing equations \eqref{eq:continuity}-\eqref{eq:temperature} are solved using SP-Wind, an in-house research code developed over the past 15~years at KU Leuven \citep[for more details, see e.g.][]{Meyers2007, Calaf2010, Munters2016, Allaerts2016, Lanzilao2024}. The spatial discretization involves a Fourier pseudo-spectral method in the horizontal directions and an energy-preserving fourth-order finite difference scheme in the vertical direction \citep{Verstappen2003}. The time integration is performed using a four-stage Runge--Kutta scheme, where the timestep is based on Courant-Friedrichs-Lewy (CFL) number of 0.4. The subgrid-scale tensors are modelled using the stability-dependent Smagorinsky model proposed by \citet{Stevens2000} with the Smagorinsky coefficient set to $C_s =0.14$ and a wall-damping function for the mixing length \citep{mason1992}, similar to the above-mentioned previous studies with SP-Wind. Turbine forces are modelled trough the widely used non-rotating actuator disk model \citep{Calaf2010, Allaerts2015}. 

\subsection{Boundary conditions}
At the top of the domain, `rigid lid' boundary conditions are adopted, i.e. the vertical derivatives of $u, v$ and $\theta$ are 0, while a Dirichlet condition is employed for vertical velocity ($w=0$). Physically, the upper boundary can be regarded as an `infinitely strong' inversion layer. The effects of the bottom surface on the flow are modelled through classic Monin--Obukhov Similarity Theory (MOST). According to MOST, the wall stress $\tau_w = \rho_0 u_*^2$ and heat flux $q_0 = -\theta_*u_*$ can be estimated from the resolved horizontal velocity $M_1 = (u_1^2 + v_1^2)^{1/2}$ and temperature $\theta_1$  at the first grid level (at height $z_1$):
\begin{align}
	u_* &= \frac{\kappa M_1}{\ln(z_1/z_{0m}) - \Psi_m(z_1/L_O) + \Psi_m(z_{0m}/L_O)} \quad \mathrm{and} \quad \\
	\theta_* &= \frac{\kappa (\theta_1-\theta_s) }{\ln(z_1/z_{0h}) - \Psi_h(z_1/L_O) + \Psi_h(z_{0h}/L_O)},
	\label{eq:MOST}
\end{align}
where $L_O = u_*^2\theta_0/(\kappa g \theta_*)$ is the Obukhov length, $\kappa=0.4$ is the Von Karman constant and $z_{0m/h}$ denote the roughness lengths for momentum and heat, respectively. In order to model the effect of the solar arrays at the ground surface, the surface temperature $\theta_s = \theta_s(x,y)$ is given as an input to the simulations, introducing spatial patterns with (high) FPV temperature $\theta_{\mathrm{FPV}}$ and (low) sea temperature $\theta_{\mathrm{sea}}$. It should be noted that the surface temperature transitions from $\theta_{\mathrm{FPV}}$ to $\theta_{\mathrm{sea}}$ are smoothed over a number of grid points using a Gaussian function, in order to avoid oscillations related to the spectral discretization \citep{Bon2023}. A uniform surface roughness of $z_0=10^{-4}$ m is assumed in all simulations here, following previous LES studies involving offshore wind farms \citep{Allaerts2017,Lanzilao2024}. Hence, the solar panels are assumed to lie directly at the sea surface and not affect surface roughness. This simplification is further discussed in section \ref{sec:discussion}. The stability correction functions $\Psi_{m/h}$ depend on the local stability conditions. For unstable surface stratification ($\theta_1 < \theta_s$, as can be expected above the solar arrays), we adopt the expressions from \citet{paulson1970}, while for stable stratification ($\theta_1 > \theta_s$) the Businger--Dyer functions are implemented \citep{Businger1971}. For more details on the stability functions and how the coupled equations in \eqref{eq:MOST} are solved, we refer to \citet{Allaerts2016} and \citet{Allaerts2018}. 

In the horizontal directions, the pseudo-spectral discretization naturally implies periodic boundary conditions. To avoid recycling of the secondary flows induced by the solar arrays and the wake caused by the wind farm, a fringe region with a concurrent precursor simulation is used \citep{Stevens2014, Munters2016, Lanzilao2022}. That is, the simulations are ran simultaneously on two domains, where the main domain contains the wind and/or solar farm while the precursor domain is `empty'. Inside the fringe region, located in the outflow of the main domain, a body force that nudges the flow fields towards the precursor field is added. Additionally, in order to avoid spanwise locking of large-scale turbulent structures that leads to undesired time-averaged high- and low velocity streaks in the precursor domain, the `shifted periodic boundary condition' approach proposed by \citet{Munters2016a} is employed. 

\subsection{Wind--Solar farm and domain configuration}
\label{subsec:configuration}
The studied wind turbines are based on the DTU 10 MW reference turbine \citep{Bak2013}, with a rotor diameter of $D=178.3$ m at hub height $z_h = 119$ m. At the wind-speeds in this study (4.7-9.5 m s$^{-1}$ at $z_h$, see section \ref{subsec:suite}), the turbines operate between the respective cut-in and rated wind speed of 4 and 11.4 m s$^{-1}$. We use a disk-based thrust coefficient of $C_T'=2$, in accordance with the Betz limit. In the considered `base configuration', the wind farm contains 45 turbines, placed in an aligned arrangement of 15 rows (streamwise) and 3 columns (spanwise). Exceptions from this base configuration are discussed in the next subsection. The spacing between the turbines is set to $s_x = s_y = 1$ km ($\approx 5.6D$), which is comparable to the operational setup of offshore wind farms Lilligrund, Sweden ($3.3-4.3D$) and Horns Rev I, Denmark ($7D$) \citep{Barthelmie2010}. 

To determine the size of the FPV arrays, we follow \citet{Golroodbari2021} in assuming that the energy density is about 100-200 MWp/km$^2$. Moreover, the power capacity of the solar farm should be comparable to the 450 MW total rated power of the wind farm, such that the grid connection can be shared \citep{Golroodbari2021, Delbeke2023}. Therefore, the total area of investigated solar farm is 1.92 km$^2$ in most simulations, corresponding to a rated power of about 192-384 MWp. In the base configuration, the solar farm consists of two strips of $l_x \times l_y = 2400 \,\mathrm{m} \times 400\, \mathrm{m}$, separated by 600 m in the spanwise direction in accordance with the turbine spacing of 1 km (\red{see also Fig. \ref{fig:secondaryflows_NoWF}a}).

The computational domain has dimensions $L_x \times L_y = 33.6 \times 6$ km in the streamwise- and spanwise directions. The domain height is fixed at $H=800$ m, which is a typical height for a neutral offshore ABL \citep{Calaf2010, Lanzilao2024}. The last 6.25\% of the streamwise direction are occupied by the fringe region ($L_{f} = 2.1$ km). The solar and/or wind farm is centred along the spanwise dimension of the domain, and starts at $x_0=4.8$ km in the streamwise direction, resulting in a downstream distance of 16.7 km from the end of the wind farm to the fringe region. As the main focus of the present study is on phenomena inside the solar/wind farm, upwind (blockage) and downwind (wake) effects are less important. Hence, considering previously used domain sizes in LES of wind farms that did investigate these effects \citep[e.g.][]{Allaerts2017, Strickland2020, Lanzilao2024}, the present domain size should be sufficient.

A grid size of $\Delta_x \times \Delta_y \times \Delta_z = 60 \times 30 \times 10$ m is employed, comparable to the LES of heterogeneous unstable channel flow from \citet{Salesky2022}, who used a uniform grid with $\Delta = 41.3$ m. The sensitivity of main results to grid resolution is further addressed in Appendix \ref{app:resolution}. We point out that the rather narrow wind farm, relatively small domain size, and coarse grid resolution employed here limit the computational cost, allowing to investigate a wide parameter space in this study.

\subsection{Suite of LES cases}
\label{subsec:suite}
An overview of the LES cases that are analyzed in this work is provided in Table \ref{tab:simulations}. The simulations are categorized into six sets or groups, labeled A through F. Set A contains simulations with only FPV arrays, all in the same base arrangement (two strips of $l_x \times l_y =2400 \mathrm{m} \times 400 \mathrm{m}$), without the inclusion of a wind farm. This layout is referred to as S2Y4X24 throughout this text, where the number behind `S' indicates the number of FPV arrays, while `Y' and `X' specify the strip width and length in hm, respectively. To investigate the formation of secondary flows under different atmospheric conditions, the driving pressure gradient, hence wind speed, and temperature difference between FPV arrays and sea surface are varied in simulation set A.

The values of the applied pressure gradient are $\{2.6, 5.2, 10.5\} \times 10^{-5}$ m s$^{-2}$, resulting in hub-height velocities of $\{4.7, 6.8,9.5\}$ m s$^{-1}$ in the precursor simulation. 
Determining the exact cell temperature of FPV modules is challenging, as it depends on various factors, including water temperature and incoming solar radiation. Moreover, there is a two-way coupling between FPV arrays and the ABL, as the cell temperature is additionally influenced by incoming wind speed and direction  \citep{Ramanan2024}. For simplicity however, we only consider fixed temperature differences ($\Delta\theta$) between the surface of the sea and the FPV modules.   \red{Both experimental and simulation studies have shown that the average surface temperature of FPV modules under solar radiation can reach up to 60$^{\circ}$C \citep{Azmi2013, Nisar2022, Ramanan2024}, while the average sea surface temperature varies between roughly 5 $^{\circ}$C at high latitudes and 30 $^{\circ}$C in the tropics \citep{Deser2010}. Hence, in order to cover a range of different realistic conditions, values of $\Delta\theta = \{5, 10, 20, 40\}$ K are investigated.}

As pointed out by previous studies that examined the effect of heterogeneous surface temperature on the ABL, the ratio between buoyancy forces and inertial forces is an important parameter that characterizes the flow \citep{Margairaz2020, Salesky2022, Fogarty2023}. Through dimensional analysis of the present set-up, we obtain a `heterogeneity Richardson number':
\begin{equation}
	Ri_h =  \frac{\frac{g}{\theta_0} \Delta\theta}{\frac{\mathrm{d} p_{\infty}}{\mathrm{d} x}} = \frac{ \frac{g}{\theta_0}\Delta\theta H}{ u_\tau^2} =  \frac{W_b^2}{u_\tau^2}
\end{equation}
with $W_b$ a buoyancy velocity scale \citep{Fogarty2023} and $u_\tau$ the friction velocity, which is equal to $(H \mathrm{d} p_\infty/\mathrm{d} x)^{1/2}$ in a channel flow. Hence, simulation set A allows to investigate the effect of $Ri_h$ through different combinations of pressure and temperature forcing (section \ref{subsec:secondaryflows}). Different forcing combinations are indicated by `U', followed by the (rounded) hub-height wind speed, and `T' followed by the value of $\Delta\theta$ (see also Table \ref{tab:simulations}). The reference Richardson number $Ri_{h,ref}=6.3\times10^3$, corresponds to cases with forcing U7T10, and is used to obtain the relative Richardson number in Table \ref{tab:simulations}. 

Subsequently, simulation set B contains simulations of the \red{base} wind farm ($3\times 15$ turbines, section \ref{subsec:configuration}) operating under the three different driving pressure gradients. The underlying surface temperature is homogeneous, and these simulations are used as reference for comparison to cases with FPV arrays.

Simulations in group C contain both wind turbines and FPV arrays. The layout of the hybrid wind--solar farm is equal in all cases, while the forcing is varied analogously to set A. These cases are discussed in section \ref{subsubsec:WF_Ri}

Furthermore, in set D, different arrangement of the solar farm are considered under equal forcing conditions. In case D1, the FPV modules are clustered in one large square array with the same total area as in group C. In cases D2-D4, the number of strips, strip width and strip length are varied. Note that when the strip width is changed, their location remains centered between the wind turbines (see section \ref{subsubsec:SF_Layout} and Fig. \ref{fig:SMs_layouts}).

In order to explore the impact of of non-parallel wind direction relative to the FPV strips, set E includes two simulations where the wind--solar farm arrangement is identical but rotated by 90 degrees (section \ref{subsubsec:wind_direction} and Fig. \ref{fig:SMs_wind_direction}). We remark that the default 3 $\times$ 15 wind farm here is replaced by a square $6\times6$ layout to limit the required domain width. Consequently, the horizontal domain size used in goup E, $L_x = L_y= 18$ km, differs from the other sets, while the grid resolution is held equal. The FPV modules are arranged in five strips of 2400 m by 200 m, resulting in a total area of 2.4 km$^2$.  Simulation E0 serves as homogeneous reference case where no solar farm is present.

Lastly, simulation set F is intended to investigate the resolution sensitivity of the results presented in this paper. Therefore, simulations F0 and F1 are identical to B2 and C3, respectively, except that the grid spacing is reduced by a factor 1.5 in all directions to $40\times 20 \times 6.7$ m$^3$. These simulations are discussed in Appendix \ref{app:resolution}.
\begingroup
\squeezetable
\begin{table}[H]
	\caption{Overview of the the LES set-up and naming used in this study. `Case' refers to the simulation set (A-F) and number, `Forcing' and `SF Layout' to the abbreviations that are used throughout the present study. The numbers behind U are the (rounded) hub-height wind speed in m s$^{-1}$, T the surface temperature difference between FPV arrays and the sea in Kelvin, S the number of FPV strips, Y and X the width and length of the strips in hm. Other columns explicitly indicate the temperature difference $\Delta\theta$, driving pressure gradient $\mathrm{d}p_\infty/\mathrm{d}x$, number of strips $N_s$, strip dimensions $l_x \times l_y$, heterogeneity Richardson number normalized by the reference $Ri_{h,ref} = 6.3 \times 10^3$ and wind-farm layout (columns $\times$ rows). The last column represents the farm-averaged power output normalized by the reference density $\rho_0$. }
	\label{tab:simulations}
	\begin{ruledtabular}
	\begin{tabular}{llcccccccc}
		Case & Forcing & SF Layout & $\Delta\theta$ [K] & $\frac{\mathrm{d}p_\infty}{\mathrm{d}x}$  [m s$^{-2}$]  & $N_s$ & $l_x\times l_y$ [hm] & $Ri_h/Ri_{h,ref}$ & WF Layout &  $\langle P\rangle$ [m$^5$ s$^{-3}$] \\
		A1   & U9T05   & S2Y4X24   & 5  & $10.4\times 10^{-5}$ & 2      & $24\times4$    & 1/4        & -    & -     \\
		A2   & U7T05   & S2Y4X24   & 5  & $5.2\times 10^{-5}$  & 2      & $24\times4$    & 1/2        & -    & -    \\
		A3   & U7T10   & S2Y4X24   & 10 & $5.2\times 10^{-5}$  & 2      & $24\times4$     & 1          & -    & -    \\
		A4   & U5T10   & S2Y4X24   & 10 & $2.6\times 10^{-5}$  & 2      & $24\times4$    & 2          & -    & -    \\
		A5   & U7T20   & S2Y4X24   & 20 & $5.2\times 10^{-5}$  & 2      & $24\times4$     & 2          & -    & -    \\
		A6   & U5T40   & S2Y4X24   & 40 & $2.6\times 10^{-5}$  & 2      & $24\times4$     & 8          & -    & -    \\
		B1   & U5-Hom  & -         & -  & $2.6\times 10^{-5}$  & -      & -       & -      			& 3$\times$15   & $5.21\times 10^5$   \\
		B2   & U7-Hom  & -         & -  & $5.2\times 10^{-5}$  & -      & -       & -          		& 3$\times$15  & $1.46\times 10^6$   \\
		B3   & U9-Hom  & -         & -  & $2.6\times 10^{-5}$ & -      & -       & -          		& 3$\times$15  &  $4.13\times10^6$  \\
		C1   & U9T05   & S2Y4X24   & 5  & $10.4\times 10^{-5}$ & 2      & $24\times4$     & 1/4        & 3$\times$15  & $4.21\times 10^6$  \\
		C2   & U7T05   & S2Y4X24   & 5  & $5.2\times 10^{-5}$  & 2      & $24\times4$     & 1/2        & 3$\times$15  & $1.54\times10^6$  \\
		C3   & U7T10   & S2Y4X24   & 10 & $5.2\times 10^{-5}$  & 2      & $24\times4$     & 1          & 3$\times$15  & $1.66\times10^6$   \\
		C4   & U5T10   & S2Y4X24   & 10 & $2.6\times 10^{-5}$  & 2      & $24\times4$     & 2          & 3$\times$15  & $6.40\times10^5$   \\
		C5   & U7T20   & S2Y4X24   & 20 & $5.2\times 10^{-5}$  & 2      & $24\times4$     & 2          & 3$\times$15  & $1.80\times10^6$   \\
		C6   & U5T40   & S2Y4X24   & 40 & $2.6\times 10^{-5}$  & 2      & $24\times4$    & 8           & 3$\times$15  & $6.85\times10^5$  \\
		D1   & U7T10   & S1Y14X14  & 10 & $5.2\times 10^{-5}$  & 1      & $14\times14$   & 1          & 3$\times$15  & $1.55\times10^6$   \\
		D2   & U7T10   & S4Y2X24   & 10 & $5.2\times 10^{-5}$  & 4      & $24\times2$    & 1          & 3$\times$15  & $1.62\times10^6$   \\
		D3   & U7T10   & S4Y4X24   & 10 & $5.2\times 10^{-5}$  & 4      & $24\times4$    & 1          & 3$\times$15  & $1.73\times10^6$   \\
		D4   & U7T10   & S2Y4X48   & 10 & $5.2\times 10^{-5}$  & 2      & $48\times4$    & 1          & 3$\times$15  & $1.79\times10^6$   \\
		E0   & U7-Hom  & -         & -  & $5.2\times 10^{-5}$  & -      & -       & -                 & 6$\times$6   &  $1.54\times10^6$  \\
		E1   & U7T10   & S5Y2X24   & 10 & $5.2\times 10^{-5}$  & 5      & $24\times2$    & 1          & 6$\times$6   &  $1.57\times10^6$  \\
		E2   & U7T10   & S5Y24X2   & 10 & $5.2\times 10^{-5}$  & 5      & $2\times24$    & 1          & 6$\times$6   &  $1.54\times10^6$  \\
		F0   & U7-Hom  & -         & -  & $5.2\times 10^{-5}$  & -      & -       & -          		  & 3$\times$15   & $1.31\times10^6$   \\
		F1   & U7T10   & S2Y4X24   & 10 & $5.2\times 10^{-5}$  & 2      & $25\times4$    & 1          & 3$\times$15   & $1.51\times10^6$  \\
	\end{tabular}
\end{ruledtabular}
\end{table}
\endgroup

\subsection{Simulation procedure}
The employed simulation practice is similar to that in previous LES studies of wind farms \citep[e.g.][]{Calaf2010, Lanzilao2024}. First, a homogeneous startup simulation (without wind- or solar farm) is performed for the three different pressure gradients, initialized from a logarithmic profile with random turbulent perturbations and run for $3 \times 10^5$ s to obtain fully developed turbulent conditions. The instantaneous flow fields of these simulations are subsequently used to initialize the corresponding concurrent precursor as well as the main domain of the simulations in Table \ref{tab:simulations}. \red{For the cases involving wind farms (sets B-F), an additional spin-up is performed to let the flow in the main domain adjust to the presence of the wind farm. The duration of this spin-up phase is set to $3\times10^4$ s ($\approx 8$ hours), corresponding to approximately 10 to 19 wind-farm flow-through times (where one flow-through time is $L_{wf}/u_{h}$). Timeseries of e.g. turbine power, $u_\tau$ and turbulent kinetic energy (not shown) reveal that this is amply sufficient to reach a statistically steady state. In the cases that involve a solar farm, the introduction of the heterogeneous surface temperature boundary conditions leads to a short transient of about 1 hour in which the thermal effects emerge. Since the length of this stage is negligible compared to the full averaging time, no additional spin-up phase was used after introduction of the solar farm. Finally, all simulations were run for a period of $8.6\times 10^5$ s ($\approx 24$ h) over which statistics were collected. Hence, all results presented below are averaged over a period of approximately one day.}




\section{Simulation results}

\subsection{Secondary flows induced by solar arrays}
\label{subsec:secondaryflows}
\begin{figure}[h!]
	\centering
	\includegraphics[width=.9\textwidth]{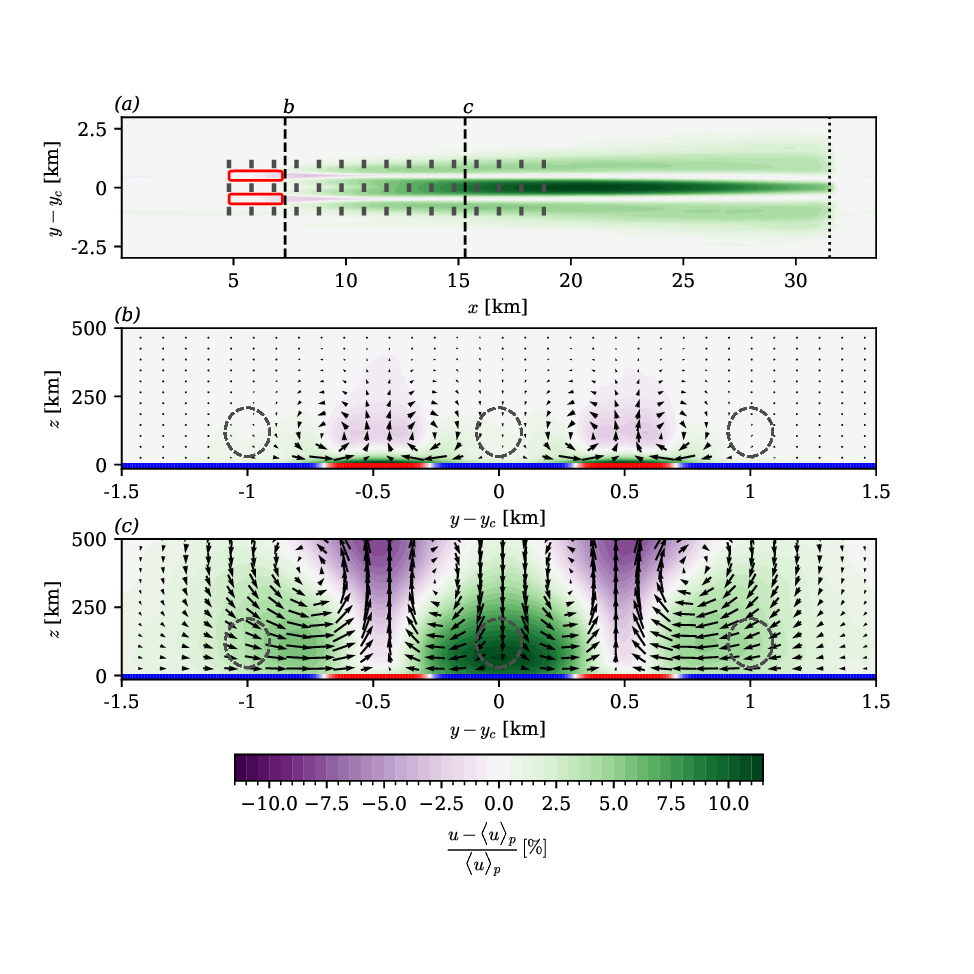}
	\caption{Time-averaged flow fields for case A3 (U7T10-S2Y4X24): contours display the local velocity difference relative to the free-stream (precursor domain) in slices at hub height (a), at $x=7.3$ km (b) and $x=15.3$ km (c). The locations of the heated solar arrays are indicated by the red rectangles in (a) and by the red colors at $z=0$ in (b,c). Vertical dashed lines show the location of planes (b,c) and the vertical dotted line marks the start of the fringe region. Vectors in (b,c) are obtained form the the time-averaged in-plane velocity components $v$ and $w$. The intended location of wind turbines, which were not present in this simulation, is indicated by vertical thick lines in (a) and dashed circles in (b,c). The $y-$axis is centered around the domain center $y_c = L_y/2$.}
	\label{fig:secondaryflows_NoWF}
\end{figure}

We first investigate the simulations with only FPV arrays, where no wind farm is present (set A). Fig. \ref{fig:secondaryflows_NoWF} shows flow fields for case A3 (U7T10-S2Y4X24). Only one case is shown here for brevity, but the patterns are similar in all simulations of group A. Panel (a) displays the relative streamwise velocity with respect to the precursor simulation at hub-height, where the location of the solar arrays in indicated by the red rectangles. Note that the potential location of wind turbines is also indicated, though they are not present in the simulations. The figure reveals a clear velocity increase at the location of the turbines, where the maximum increase of about 10\% is located in the center of the domain, approximately 10 km downstream of the solar arrays. Panels (b) and (c) confirm that the horizontal velocity variations are associated with secondary vortices, producing low-momentum pathways above the FPV arrays. In panel (b), located at the end of the solar arrays, the secondary flows are seen to be generated at the edge of the high-temperature patches and start relatively weak. At $x=15.3$ km, the strength of the vortices has increased and the center has moved upwards, in accordance with the findings in \citet{Bon2023}. There however, the secondary flows were damped by stable stratification, while in the present simulations the vortex centers keep rising in the downstream direction, until the top of the boundary layer is reached. We note that the local heating by FPV arrays at the bottom surface does not significantly alter the mean neutral stratification of the boundary layer, as their area is relatively small compared to the domain size.

\FloatBarrier

\begin{figure}[h!]
	\centering
	\includegraphics[width=.9\textwidth]{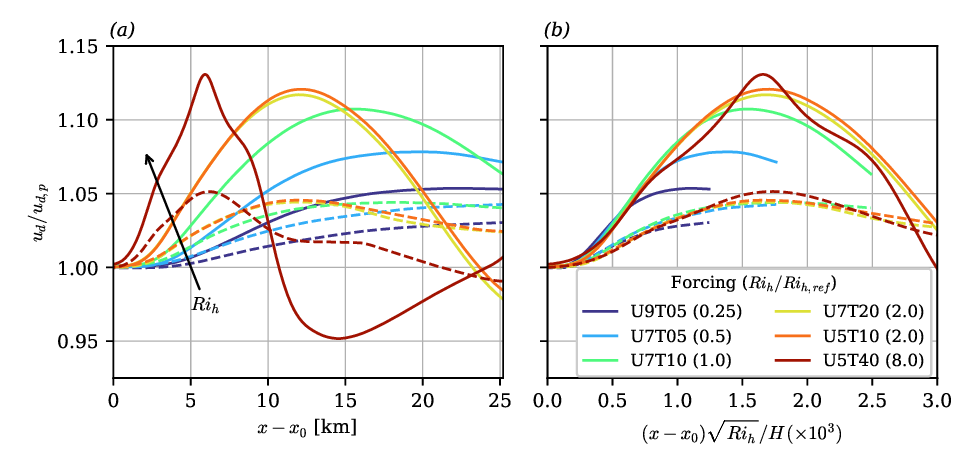}
	\caption{Streamwise development of axial velocity for simulations of set A, averaged over potential wind turbine columns in the flow field (see Eq. \eqref{eq:ud}), normalized with the free-stream value obtained from the precursor simulation. All cases have the same FPV layout (S2Y4X24) and do not contain actual wind turbines, while the forcing (pressure gradient and FPV temperature) are varied. Lines are colored by increasing Richardson number, where full lines represent the center turbine column and dashed lines are averaged of the two outer columns.}
	\label{fig:ux}
\end{figure}

To compare the velocity increase for cases with different forcing, we consider the disk-averaged velocity:
\begin{equation}
	\red{u_d(x)~=~\frac{4}{\pi D^2}\int \int_{A_d} u(x,y,z) \,\mathrm{d}A_d,}
	\label{eq:ud}
\end{equation}
with $A_d$ the area of the rotor disc. This quantity, normalized by the free-stream value from the precursor simulation, $u_{d,p}$, is presented in Fig. \ref{fig:ux}, where the fringe region is excluded. Since the cases are sorted by the Richardson number, it is clear that a higher $Ri_h$ causes a larger increase that occurs closer to the FPV arrays. The behaviour of cases U7T20 and U5T10 is approximately identical, as they have equal Richardson numbers. We note that there is a velocity deficit after $x-x_0 \approx 10$ km in the case with the most extreme $Ri_h$, where the secondary flows reach the top and side edges of the domain (not shown). Hence, the results of this case may be significantly affected by the domain size and should be interpreted carefully. Furthermore, the velocity increase in the center column (full lines) is significantly larger than in the outer columns (dashed lines), as the center is affected by the downdrafts of two secondary vortices combined (cf. Fig. \ref{fig:secondaryflows_NoWF}(b) and (c)).  

The trends in Fig. \ref{fig:ux}(a) suggest that the streamwise development of the velocity scales with $Ri_h$. Similar to \citet{Salesky2022}, who presented a scaling analysis concerning secondary motions induced by streamwise infinite strips of varying heat flux, a very basic scaling argument can be derived from the continuity equation. Ignoring spanwise variations in the streamwise development, we derive:
\begin{equation}
	0= \frac{\partial u}{\partial x} + \frac{\partial w}{\partial z} \sim \frac{u_\tau}{\mathcal{L}_x}+\frac{W_b}{H} \rightarrow \mathcal{L}_x \sim H \frac{u_\tau}{W_b} = \frac{H}{\sqrt{Ri_h}},
	\label{eq:streamwise_lengthscale}
\end{equation}
where $\sim$ indicates `scales with'.  Assuming that vertical velocity $w \sim W_b$, streamwise velocity $u \sim u_\tau$ and the vertical extent of the secondary flows scales with the boundary-layer height $H$, we find that the streamwise scaling length $\mathcal{L}_x$ scales with $H/\sqrt{Ri_h}$. In Fig. \ref{fig:ux}(b), the streamwise coordinate has been divided with this length scale, resulting in a reasonable agreement of all cases for the center column as well as the outer columns. The collapse is most convincing in the vicinity of the solar farm, whereas the lines diverge further downstream as the peak height is not scaled.

\FloatBarrier

\subsection{Thermally-induced secondary motions in hybrid wind--solar farms}
\label{subsec:sm_wsf}
\begin{figure}[h!]
	\centering
	\includegraphics[width=.9\textwidth]{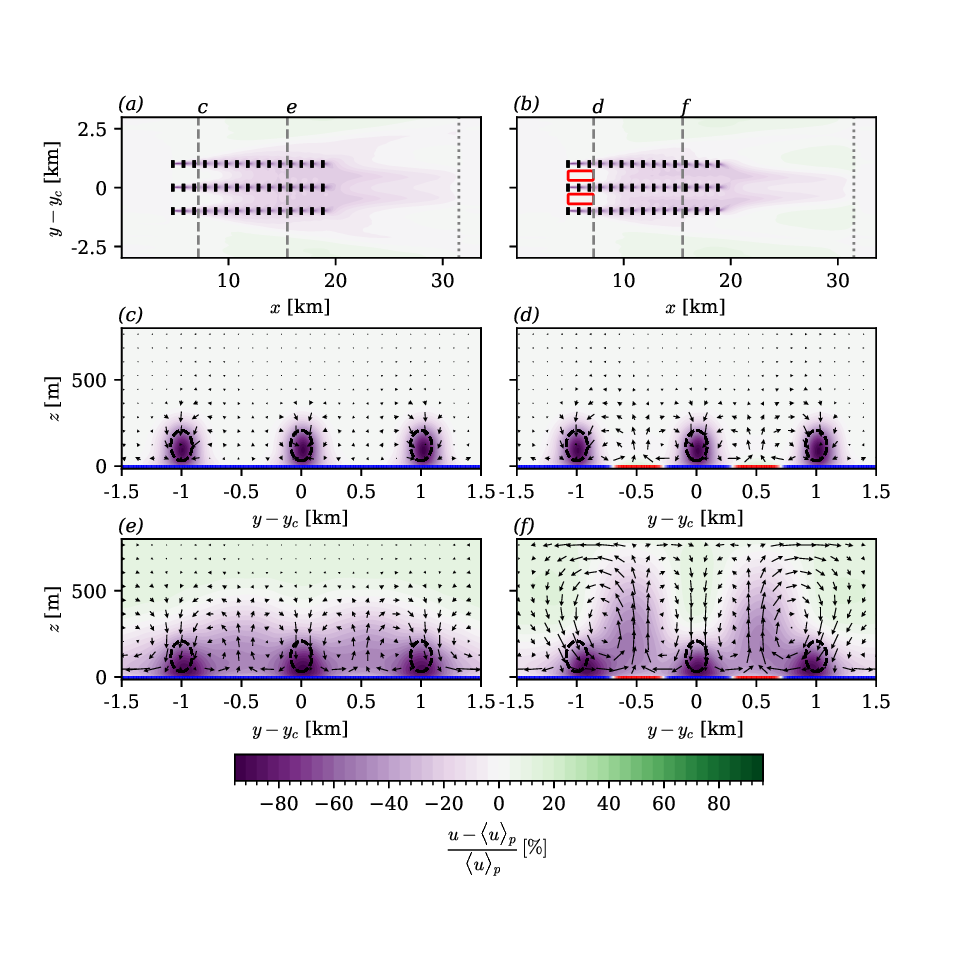}
	\caption{Time-averaged flow fields in a wind farm without (left, case B2 (U7-Hom)) and with FPV arrays (right, case C3 (U7T10-S2Y4X24): contours display the local velocity difference relative to the free-stream (precursor domain) in slices at hub height (a,b), at $x=7.3$ km (c,d) and $x=15.3$ km (e,f). The locations of the warmer solar arrays are indicated by the red rectangles in (b) and by the red colors at $z=0$ in (d,f). Vertical dashed lines in (a,b) show the location of planes (c-f) and the vertical dotted line marks the start of the fringe region. Vectors in (b,c) are obtained form the time-averaged in-plane velocity components $v$ and $w$. The location of wind turbines is indicated by vertical thick lines in (a,b) and dashed circles in (c-f).}
	\label{fig:comparison}
\end{figure}

Next, we consider how the secondary flows that are induced by the FPV arrays affect the power generation of a wind farm. Figure \ref{fig:comparison} provides a comparison between a wind farm without (left panels) and with FPV strips (right panels) operating under the same pressure gradient ($u_{h} = 6.7$ m s$^{-1}$). 
Figure \ref{fig:comparison}(c) shows strong velocity deficits in the wakes downstream of the wind turbines, due to their energy extraction. Deeper in the wind farm (Fig. \ref{fig:comparison}(e)), mean streamwise vortices, which resemble the thermally-induced secondary flows from the previous section, can be observed, leading to entrainment of high-momentum fluid into the turbine wake as previously shown by e.g. \citet{Meyers2013a} and \citet{Hodgkin2023}.

Figures \ref{fig:comparison}(d) and (f) illustrate that these circulations are considerably stronger with the inclusion of the solar farm. Hence, the already present `turbine-induced' secondary motions are reinforced by the thermally-induced secondary flows, enhancing vertical mixing throughout the entire boundary layer. The augmented downward entrainment of vertical momentum at the turbine locations can be expected to promote wake recovery and enhance overall wind-farm efficiency.

\FloatBarrier

\subsubsection{Effect of pressure and temperature forcing}
\label{subsubsec:WF_Ri}
\begin{figure}[h!]
	\centering
	\includegraphics[width=.9\textwidth]{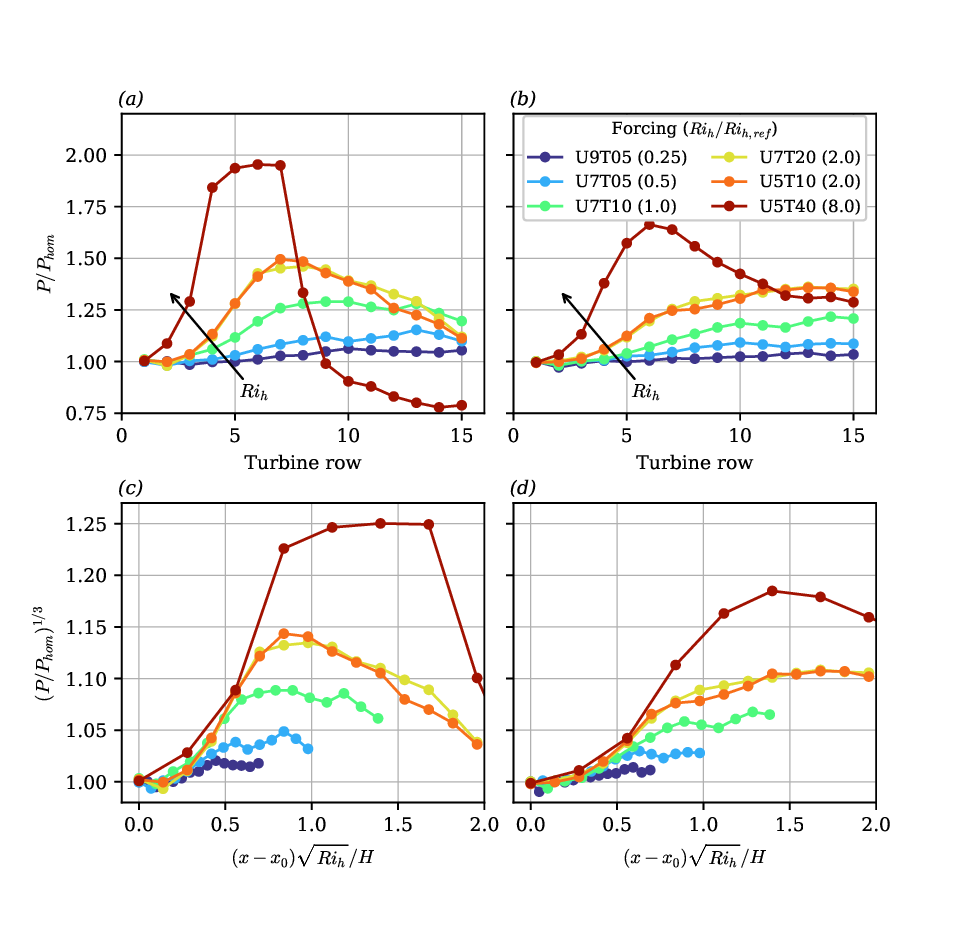}
	\caption{Time-averaged turbine power increase in solar-wind farms compared to a homogeneous wind farm for center column (a,c) and outer columns (b, d). In panels (c,d), the $x$-axis is scaled with the characteristic streamwise length scale $\mathcal{L}_x$ (Eq. \eqref{eq:streamwise_lengthscale}). Only cases of set C, with the same wind--solar farm layout (S2Y4X24) but different pressure and temperature forcing, are shown. Lines are colored by forcing and sorted by Richardson number, similar to Fig. \ref{fig:ux}. }
	\label{fig:Px_Forcing}
\end{figure}

\begin{figure}[h!]
	\centering
	\includegraphics[width=.9\textwidth]{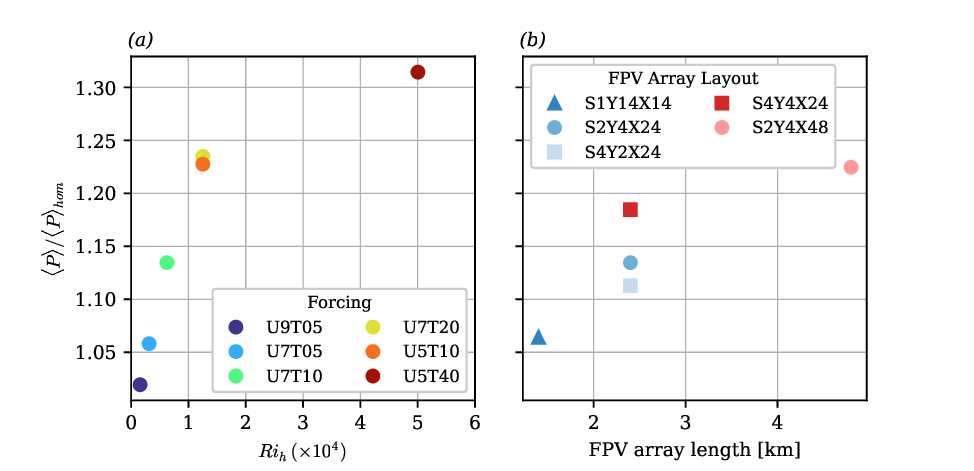}
	\caption{Farm-averaged power in hybrid solar-wind farms compared to a homogeneous wind farm with the same pressure gradient. Panel (a) shows cases with different pressure and temperature forcing of set C as function of the Richardson number, with colors equal to Fig. \ref{fig:Px_Forcing}. Panel (b) displays cases with different solar farm layout and identical pressure and temperature forcing of set D (and case C3), as a function of the length of the FPV strips. Simulations with doubled total FPV area are represented by red markers, while cases with equal number of strips are indicated by identical marker shapes.}
	\label{fig:P_Ri_lengt}
\end{figure}

The effect of the secondary flows, generated by the FPV arrays, on the wind-farm power generation is quantified in Figs. \ref{fig:Px_Forcing} and \ref{fig:P_Ri_lengt}. Figure \ref{fig:Px_Forcing} compares the power per turbine of the center column (a) and outer columns (b) to the simulations of a homogeneous wind farm operating under the same pressure gradient. Consistent with the velocity changes in the previous section, the power increase depends on the Richardson number, and is small at the farm entrance but grows further downstream. Further comparing to Fig. \ref{fig:ux}, we note that the difference in power increase between the outer and center turbines (panel a versus panel b) is smaller than the difference between the velocity increase at the corresponding locations in the cases without wind farm.

Analogous to the apparent scaling of the streamwise velocity development found in the previous section, the streamwise location of the turbines is scaled with the  characteristic length scale $\mathcal{L}_x = H/\sqrt{Ri_h}$ in Figs. \ref{fig:Px_Forcing}(c) and (d). The exponent in the relative power increase $(P/P_{hom})^{1/3}$ is motivated by the fact that $P \sim u^3$. The collapse of the curves for cases with different forcing is reasonable near the farm entrance, but less convincing than in Fig. \ref{fig:ux}(b). 

The farm-averaged performance is compared in Fig. \ref{fig:P_Ri_lengt}(a), revealing that the total power improvement due to the FPV arrays rises for increasing Richardson number, i.e. for high FPV temperatures in combination with low wind speeds. Even though the power in the last center rows of case U5T40 is actually lower than the homogeneous case, the total power increase is highest with more than 30\%.

\FloatBarrier

\FloatBarrier

\subsubsection{Effect of solar farm arrangement}
\label{subsubsec:SF_Layout}
\begin{figure}[h!]
	\centering
	\includegraphics[width=.9\textwidth]{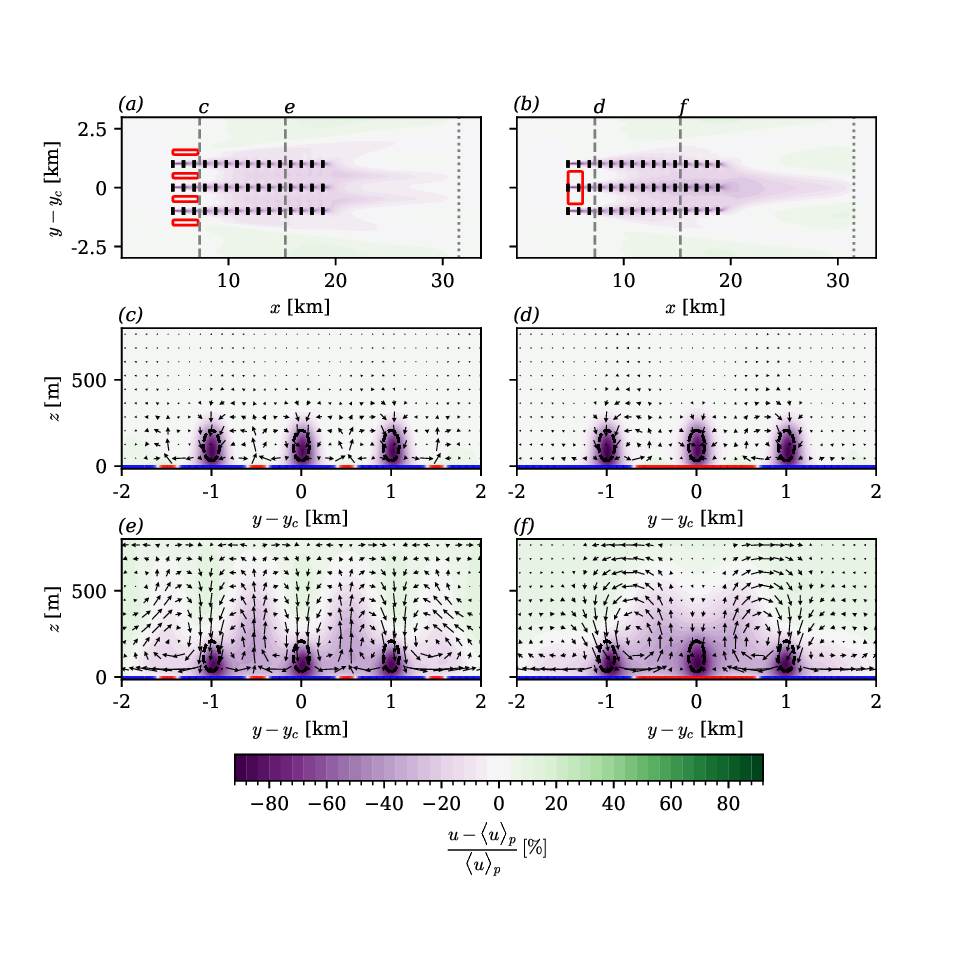}
	\caption{Time-averaged flow fields in two wind--solar farms with different FPV configuration: case D3 (layout S4Y4X24, left) and case D1 (layout S1Y14X14, right). Contours display the local velocity difference relative to the free-stream (precursor domain) in slices at hub height (a,b), at $x=7.3$ km (c,d) and $x=15.3$ km (e,f). The locations of the warmer solar arrays are indicated by the red rectangles in (a,b) and by the red colors at $z=0$ in (c-f). Vertical dashed lines in (a,b) show the location of planes (c-f) and the vertical dotted line marks the start of the fringe region. Vectors in (b,c) are obtained form the the time-averaged in-plane velocity components $v$ and $w$. The location of wind turbines is indicated by vertical thick lines in (a,b) and dashed circles in (c-f).}
	\label{fig:SMs_layouts}
\end{figure}

\begin{figure}[h!]
	\centering
	\includegraphics[width=.9\textwidth]{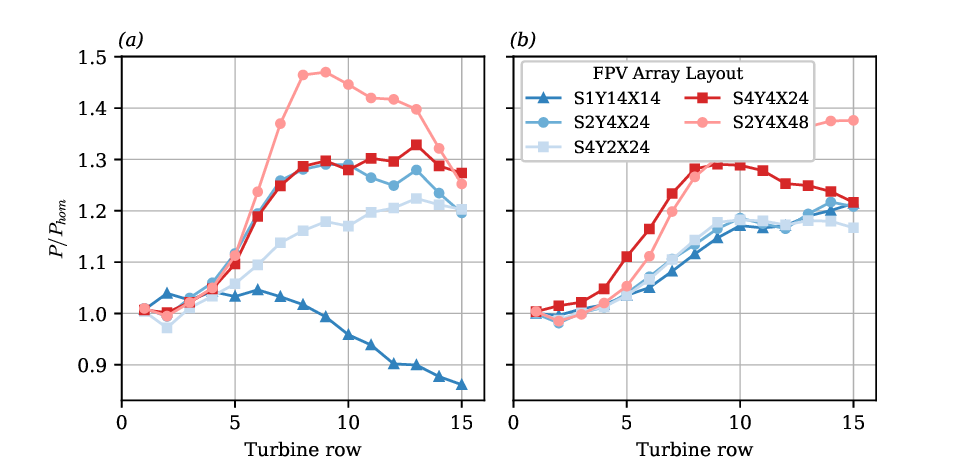}
	\caption{Time-averaged turbine power increase in solar-wind farms with different FPV configuration, compared to a homogeneous wind farm for center column (a) and outer columns (b). Only cases from set D and case C3 are shown, which have identical pressure and temperature gradient. Cases with doubled total FPV area are represented in red, while cases with equal number of strips are indicated by identical markers.}
	\label{fig:Px_Layout}
\end{figure}

Next, we compare cases from set D and case C3, with equal pressure forcing and FPV temperature, but different configurations of the solar arrays. Figure \ref{fig:SMs_layouts} (a,c,d) shows time-averaged flow fields of case S4Y4X24, which is similar to the standard configuration  (S2Y4X24 in Fig. \ref{fig:comparison}) but with four smaller strips instead of two strips of FPV. The additional strips that are outside the wind farm are seen to reinforce the secondary flow pattern in the outer columns. Figure \ref{fig:Px_Layout}(b) confirms that these stronger secondary flows leads to more power enhancement in the outer columns, while Fig. \ref{fig:Px_Layout}(a) shows that the power increase in the center column is  more similar to the two-strip case S2Y4X24. The latter suggests that the secondary flows in the center column are not substantially affected by the total number of FPV strips.

The right-hand column of Fig. \ref{fig:SMs_layouts} depicts the flow field in case D1 (layout S1Y14X14), where the FPV area is equal to our standard configuration but arranged in one square array of $1.4 \times 1.4$ km$^2$. Figures  \ref{fig:SMs_layouts}(d,f) reveal that two secondary vortices form, centered above the spanwise edges of the FPV array. However, the high-temperature patch below the center turbines induces upward velocity through positive buoyancy forces, thereby reducing entrainment of high-momentum fluid from above. Figure \ref{fig:Px_Layout}(a) demonstrates that this results in a diminished power output of the turbines in the center column. In contrast, the two secondary vortices at the edges of the single FPV array lead to a power increase in the two outer turbine columns that is comparable to the default configuration, as shown in Fig. \ref{fig:Px_Layout}(b). 

A few more observations can be made regarding the comparison of cases with different FPV layout in Fig. \ref{fig:Px_Layout} and the total wind-farm power output in Fig. \ref{fig:P_Ri_lengt}. The lighter red line in Fig. \ref{fig:Px_Layout}, representing a case where the length of the solar strip is doubled, shows that longer FPV arrays lead to augmented power increase, especially in the second half of the wind farm. Figure \ref{fig:P_Ri_lengt}(b) suggests that in general, longer FPV arrays, as well as a larger total FPV area, improve wind-farm power generation.  Comparing cases with equal FPV area (blue), we find that four smaller strips (S4Y2X24) induce slightly smaller power enhancement than two wider strips (S2Y4X24). The wind--solar farm with the `naive' square FPV array produces  6\% less power. Furthermore, the two cases with doubled FPV area (red) suggest that two longer arrays (S2Y4X48) are more efficient in increasing power output than four shorter strips (S4Y4X24).

\FloatBarrier

\subsubsection{Impact of wind direction}
\label{subsubsec:wind_direction}
\begin{figure}[h!]
	\centering
	\includegraphics[width=.9\textwidth]{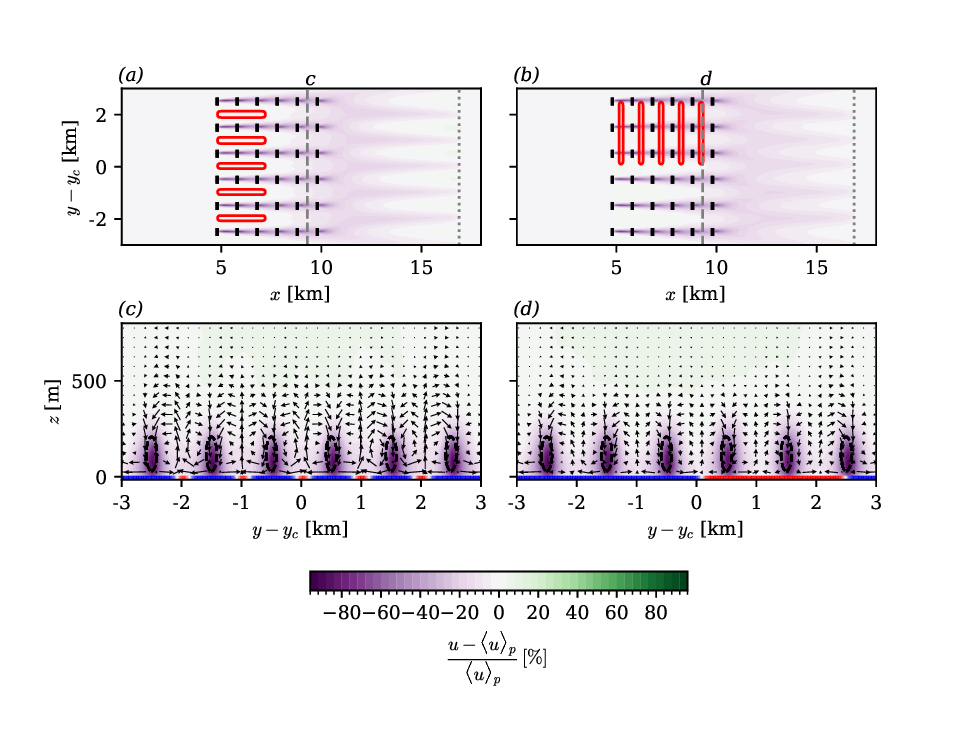}
	\caption{Time-averaged flow fields for parallel (left, case E1) and perpendicular (right, case E2) wind direction. Contours display the local velocity difference relative to the free-stream (precursor domain) in slices at hub height (a,b) and at $x=7.3$ km (c,d). The locations of the warmer solar arrays are indicated by the red rectangles in (a,b) and by the red colors at $z=0$ in (c-f). Vertical dashed lines in (a,b) show the location of planes (c,d) and the vertical dotted line marks the start of the fringe region. Vectors in (b,c) are obtained form the the time-averaged in-plane velocity components $v$ and $w$. The location of wind turbines is indicated by vertical thick lines in (a,b) and dashed circles in (c-f).}
	\label{fig:SMs_wind_direction}
\end{figure}

\begin{figure}[h!]
	\centering
	\includegraphics{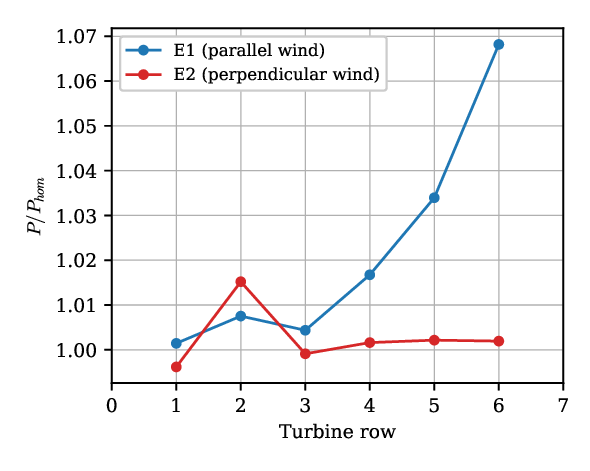}
	\caption{Time-averaged turbine power for cases with parallel (case E1, blue) and perpendicular (case E2, red) wind direction, normalized with power in a homogeneous wind farm. Powers are averaged over the entire row (i.e. 6 turbines). }
	\label{fig:Px_wind_direction}
\end{figure}

Until now, the wind direction was `ideal' for the generation of secondary flows, i.e. parallel to the FPV strips \citep{Anderson2020, Mironov2024}. Figure \ref{fig:SMs_wind_direction} shows case E1 (left) and case E2 (right), where the inflow direction is perpendicular to the FPV arrays. Secondary vortices are present in both cases, yet they are found to be weaker in the case with perpendicular inflow. Considering the row-averaged power relative to a homogeneous wind farm, as displayed in Fig. \ref{fig:Px_wind_direction}, we observe that the perpendicular FPV strips induce no substantial power increase, nor a decrease. We note that the power increase in the case with parallel inflow is significantly smaller than in the previously discussed cases (cf. Figs. \ref{fig:Px_Forcing} and \ref{fig:Px_Layout}), because the wind farm here is only 6 rows instead of 15 rows.

\FloatBarrier

\section{Discussion}
\label{sec:discussion}
Before the present insights can be applied to real-world wind--solar farms, several limitations arising from the simplifying assumptions that have been made throughout this study need to be discussed. Firstly, the neutrally stratified PDBL considered here ignores many aspect of the vertical structure in the real ABL. The height of the boundary layer is fixed and the rigid lid upper boundary conditions can be regarded as an infinitely strong thermal inversion layer, while in reality the presence of large flow perturbations such as wind farms, and perhaps even large FPV arrays, can cause vertical displacement of the capping inversion aloft, inducing atmospheric gravity waves and affecting flow blockage \citep{Allaerts2017, Lanzilao2024}.   
Previous studies have demonstrated that thermal stratification within and above the ABL strongly impacts wind-farm performance \citep[e.g.]{Allaerts2017, Allaerts2018, Strickland2022a, Maas2022, Lanzilao2024}. Moreover, the PDBL does not account for Coriolis forces, which gives rise to rotation of the wind along the vertical direction, also known as the Ekman spiral. The PDBL assumption is reasonable when wind turbines are located in the inner layer (up to $\sim 0.2H\approx 160$ m \cite{Calaf2010}), which is slightly violated for the presently considered turbines with a tip height of 205 m. Moreover, the thermally-induced secondary flows do penetrated deeper into the boundary layer and can therefore be expected to be affected by Coriolis forces and the Ekman spiral. We note however that, to the authors' best knowledge, the effect of Coriolis forces on secondary motions generated by spanwise surface heterogeneity has not been addressed yet in the literature. 

Regarding the representation of the FPV arrays, only one-way coupling between heated solar installations and the ABL has been considered, while FPV temperature is additionally affected by the incoming wind speed and solar radiation \citep{Ramanan2024}. More sophisticated LES and surface models would be required to include the interaction between FPV arrays and the ABL. Furthermore, if the FPV installation involves tilted or elevated solar panels, the surface roughness would be modified. It is known that large-scale spanwise roughness heterogeneity generates secondary flows of Prandtl's second kind \citep{Anderson2015a}, where the rotational direction appears to depend on the protruding height of the roughness \citep{Stroh2020a}, but the generation of secondary flows over combined roughness-temperature heterogeneity has not been investigated yet.


Lastly, we point out that the increased power enhancement in hybrid wind--solar farms for high Richardson numbers is particularly favourable considering the complementarity of wind and solar power and cable `pooling' \citep{Golroodbari2021} mentioned in the introduction. Large $Ri_h$ can be expected in fair weather conditions; with weak wind, clear skies and strong solar radiation leading to high FPV temperature. Consequently, the wind farm is operating well below rated power, making the higher power improvement more beneficial than in situations with high wind speeds. However, the LES cases here only cover a limited range of conditions. Investigating a more extensive range of conditions may be an interesting topic for future research. 




\section{Conclusions}
\label{sec:conclusion}
The set of LES cases in the present study provides evidence that surface temperature differences induced by floating solar installations at sea can significantly affect the surrounding ABL flow, which could be exploited to enhance the power generation of the wind turbines in a hybrid solar-wind farm. When FPV arrays are arranged in long strips ($2400\times400$ m here) parallel to the flow, the heterogeneous buoyancy force generates large-scale secondary vortices that penetrate deep into the boundary layer and produce lateral variations of mean streamwise velocity. Simulation results indicate that the ratio between the driving pressure gradient and the temperature difference between the solar panels and the sea, quantified by a heterogeneity Richardson number $Ri_h$, is an important parameter that determines the downstream development of the streamwise velocity increase in the regions between the FPV strips. When wind turbines are placed strategically in these locations, the power production of downstream turbines can be increased up to a factor two compared to the same wind farm without FPV strips, as the secondary motions enhance entrainment of high momentum from above the farm. The farm-averaged power improvement is amplified as $Ri_h$ rises, up to 30\% for the largest value considered, which occurs for high FPV temperature and low wind speeds. Comparing different arrangements of the FPV arrays, it is shown that longer strips are beneficial for the wind-farm power generation. Lastly, we demonstrate that the power production in a hybrid wind--solar farm is approximately equal to an isolated wind farm when the flow direction is perpendicular to the FPV strips.

Although these results seem promising, the present study is the first of its kind and many simplifications have been made. Thus, deeper investigations are needed to better understand and quantify the interaction between hybrid wind--solar farms and the ABL. Further research should consider a wider range of less idealized atmospheric conditions, as well as more realistic representation of FPV arrays at the surface. Finally, experimental studies using wind tunnels would be helpful for verification.

\begin{acknowledgements}
The computational resources and services used in this work were provided by the VSC (Flemish Supercomputer Center), funded by the Research Foundation – Flanders (FWO) and the Flemish Government. The authors acknowledge financial support from the Research Foundation Flanders (T.B., FWO grant G098320N). 
\end{acknowledgements}

\section*{Author contributions}
T.B. performed code implementations, worked on visualization and interpretation of results, and wrote the original draft. V.C. performed the numerical experiments, contributed to conceptualization and interpretation. J.M. supervised the research, reviewed the original draft and was responsible for acquisition of funding.

\appendix*
\section{Resolution sensitivity}
\label{app:resolution}
\begin{figure}[h!]
	\centering
	\includegraphics[width=\textwidth]{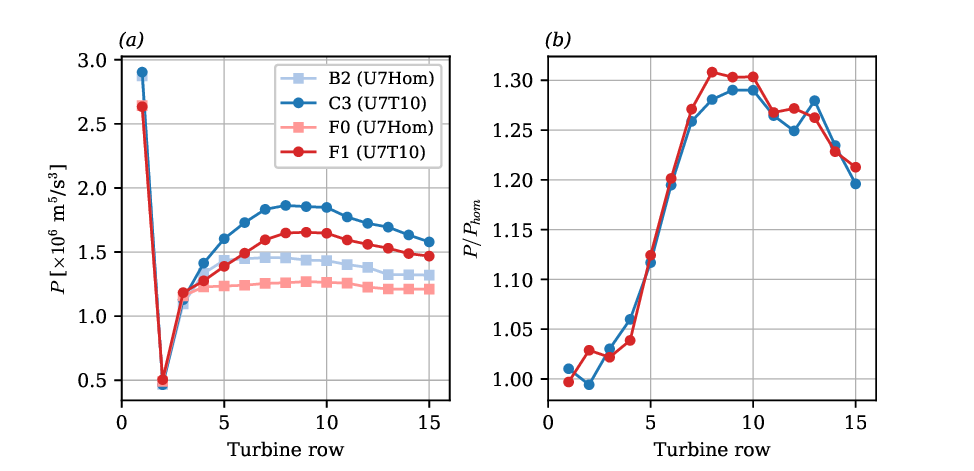}
	\caption{Time-averaged power output for turbines in the center column of homogeneous wind farms (lighter colors) and wind--solar farms (darker colors) in simulation with the default (blue) and finer (red) grid resolution. Panel (b) displays the power increase in the hybrid wind--solar farm with respect to the homogeneous wind farm at equal resolution (i.e. C3 divided by B2 and F1 divided by F0)  }
	\label{fig:Px_Resolution}
\end{figure}

In order to assess the dependency of the results in this paper on the grid spacing in of the LES, two simulations where the number of grid points was increased by a factor 1.5 in all directions where performed (set F in Table \ref{tab:simulations}). In Fig. \ref{fig:Px_Resolution}(a), we compare the wind farm power of cases F0 and F1 to the corresponding simulations at lower resolution, i.e. cases B2 and C3, respectively. It is clear that the turbine power is reduced in the simulations with higher resolution. This may be linked to the fact that the employed grid resolution of $60\times30\times10$ m leads to a total of 6 and 18 grid points along the turbine rotor disk width and height, which is slightly less than that recommended by \citet{Allaerts2017} (7 and 20) and \citet{Lanzilao2024} (9 and 40) who utilized the same SP-Wind LES code. Nevertheless, the most important results in the present paper are the comparisons between the power generation in homogeneous wind farms and hybrid wind--solar farms, which is clearly less affected by the grid resolution as shown in Fig. \ref{fig:Px_Resolution}(b). Lastly, we point out that the farm-averaged power ratio $\langle P\rangle/\langle P_{hom}\rangle$ is 1.13 for the resolution used throughout the paper and 1.15 for the finer grid of set F. 

\FloatBarrier


\bibliography{library.bib}

\bibliography{library.bib}

\end{document}